\documentclass[12pt]{article}
\usepackage{cite,epsfig,amssymb,amsmath,graphicx,color}
\usepackage{stackengine,scalerel} 
\usepackage{soul}
\usepackage{xcolor,cancel}
\evensidemargin \oddsidemargin

\newcommand{\nn}{\nonumber}

\usepackage{arydshln} 
\usepackage{multirow} 

\def\lambdabar{\ThisStyle{\ensurestackMath{\stackon[-2.4\LMpt]{%
  \SavedStyle\lambda}{\kern-.5pt\kern\LMpt\rule{1\LMex}{.25pt+.15\LMpt}}}}} 

\begin{document}

\begin{center}
{{{\Large \bf Gravitational Waves as a Probe of the  Extra Dimension}
}\\[17mm]
O-Kab Kwon$^{1}$, Seokcheon Lee$^{1}$, D. D. Tolla$^{1,2}$\\[3mm]
{\it $^{1}$Department of Physics,~BK21 Physics Research Division,
~Institute of Basic Science, Sungkyunkwan University, Suwon 16419, Korea\\
$^{2}$ University College,
Sungkyunkwan University, Suwon 16419, South Korea}\\[2mm]
{\it okab@skku.edu,~skylee2@gmail.com,~ddtolla@skku.edu} }

\end{center}
\vspace{15mm}

\begin{abstract}
We consider the Einstein-Hilbert action without cosmological constant in 5-dimensions  and implement the Kaluza-Klein (KK) reduction by compactifying the fifth direction on a circle of small but finite radius. For non-zero compactification radius, the 4-dimensional spectrum contains massless and massive KK modes.  For the massive KK modes, we retain four
KK tensor and one KK scalar modes after a gauge fixing. We treat those massive KK modes as stochastic sources of gravitational wave (GW) with characteristic dependences of the frequencies on the size of the extra dimension.  Using the observational bounds on the size of the extra dimension and on the characteristic strain, we make an order estimation on the frequencies and amplitudes of the massive KK modes that can contribute to the GW.  

\end{abstract}

\newpage

\section{Introduction}
\label{sec:intro}

The detection of gravitational waves (GWs) by the LIGO and VIRGO collaborations~\cite{Abbott:2016blz,Abbott:2016izl,Abbott:2016nmj,TheLIGOScientific:2016pea,Abbott:2017vtc,Abbott:2017oio,Abbott:2017gyy,Nitz:2018imz,LIGOScientific:2018mvr} from a merging of a pair of heavy black holes seems to be consistent with a prediction of Einstein's general relativity. 
The GW due to the merging of a neutron star pair has also been observed subsequently~\cite{TheLIGOScientific:2017qsa,GBM:2017lvd}, opening new possibilities for astronomical observation and cosmological research.
Shortly after the general relativity was established, the idea to unify the gravity and the electromagnetic force, known as the Kaluza-Klein (KK) reduction~\cite{Nordstrom1914,1921SPAW.......966K,1926ZPhy...37..895K,1981NuPhB.186..412W,Overduin:1998pn}, was born.  
This was an innovative attempt based on the existence of extra dimensions, and later the presence of extra dimensions became an essential element in the study of unified theories, such as string/M-theories.
For this reason, if one can prove the existence of extra dimensions in nature, a new horizon of understanding of physics will be opened. Despite many attempts to prove it, it has not been successful up to now.

The current theoretical prediction of the GW is based on the general relativity in 4-dimensions 
with no reference to the extra dimensions.  
However, if we are living in a Universe with extra dimensions, there can be some remnants of the extra dimensions in the detection of the GW. 
The remnants may encode some information of the extra dimensions, such as the size of the extra dimension and the dynamics of fluctuations modes. Earlier attempts to explain the effect of extra dimensions on GW can be found in \cite{Barvinsky:2003jf,Cardoso:2002pa,Cardoso:2003jf,Cardoso:2008gn, Alesci:2005rb,Yu:2016tar, Andriot:2017oaz, Chakraborty:2017qve, Visinelli:2017bny, Pardo:2018ipy}.

In this paper, we consider the KK reduction of the Einstein-Hilbert action in 5-dimensions without a cosmological constant, which has one compactified spatial dimension of small but finite size $L$.\footnote{To provide a physical interpretation for the Kaluza's cylinder condition~\cite{1921SPAW.......966K}, Klein considered the same setting in the $L\to 0$ limit~\cite{1926ZPhy...37..895K}. So the information of the extra dimension is not encoded in this reduction procedure.}
For nonzero $L$, the momenta of massive KK modes contain the information of the size of the extra dimension. 
We explicitly show that these massive modes contribute to metric fluctuations, which can have effects on the GW detections. To see the leading contribution in the small $L$ limit, we analyze the 5-dimensional Einstein equation up to quadratic order in the metric fluctuations. We briefly summarize the procedure below.

The background manifold in our setting is given by ${\cal M}_4\times S^1$, where ${\cal M}_4$ is the 4-dimensional Minkowski space.\footnote{One can naturally extend our setting to the case of the $(4+n)$-dimensional spacetime with $n$-dimensional extra dimensions with Ricci flat background. Then the background manifold becomes ${\cal M}_4\times T^n$, where $T^n$ represents the $n$-torus. Since there will be no fundamental difference with the case of the 5-dimensional setting, we concentrate on the 5-dimensional case, for simplicity.} Then the 5-dimensional metric is expressed as 
\begin{align}\label{gpq}
g_{pq}(x,y) = g^{(0)}_{pq}(x,y) + h_{pq}(x,y),
\end{align}
where $g^{(0)}_{pq}$ represents the background metric, $h_{pq}$ is the metric fluctuation, and $x,y$ are the coordinates for 4-dimensions and the fifth-dimension, respectively.  In order to investigate the effect of the extra dimension on the 4-dimensional GW, we insert \eqref{gpq} into the 5-dimensional Einstein equation and keep terms up to quadratic order in $h_{pq}$. We expand the fluctuations in terms of the spherical harmonics $Y^I(y)$ on $S^1$  as $h_{pq}(x,y) = h^I_{pq}(x) Y^I(y)$ and then project the Einstein equations on those spherical harmonics in order to obtain the equations of motion for each KK mode $h^I_{pq}$.  
With an appropriate gauge choice, we show that the set of the dynamical fields is composed of the massless graviton $\hat h^0_{\mu\nu}$, two massless vector modes $v_{\mu}^0$, one massless scalar mode $s^{0}$, the massive KK scalar modes $h^I$, and the massive KK graviton modes $\hat h^{I}_{\mu\nu}$ with $I=1,2,\cdots$.\footnote{We denote the 5-dimensional spacetime indices as $p,q, \cdots$, the 4-dimensional ones as $\mu,\nu, \cdots$,  and the index $5$ for the extra dimension.} 
The massive KK modes, $h^I$ and $\hat h^{I}_{\mu\nu}$, behave as matter fields and plays the role of the source for metric fluctuations in the equation of motion for the massless graviton $\hat h^0_{\mu\nu}$. This is given by
\begin{align}\label{msteq}
\square \bar h^0_{\mu\nu} = -\frac{16\pi G_4}{c^4} T_{\mu\nu},
\end{align}
where $G_4$ is the 4-dimensional Newton constant,  $\bar h^0_{\mu\nu}$ denotes the trace-reversed graviton mode defined as $\bar h^0_{\mu\nu}\equiv \hat h^0_{\mu\nu} - \frac12 g_{\mu\nu} \hat h^{0\rho}_{~~\rho}$ satisfying the Lorentz gauge $\nabla^\mu \bar h^0_{\mu\nu} = 0$. 
The energy-momentum tensor $T_{\mu\nu}$ is built from the massive KK modes and their derivatives. 

The form of the $T_{\mu\nu}$ in \eqref{msteq} is complicated and non-canonical in the sense that it includes {\it higher derivatives}. This makes the task of understanding the effect of the massive KK modes on the 4-dimensional GW highly non-trivial.  In order to simplify the problem, we assume that the sources of GW from the extra dimension are stochastic.  This assumption is realized by the ergodic average of $T_{\mu\nu}$ denoted by $\langle T_{\mu\nu}\rangle$~\cite{Maggiore:1999vm,Maggiore2008}. The equation of motion for the massive KK tensor modes implies that it is traceless and transverse. Due to these properties, the ergodic average of the energy-momentum tensor $\langle T_{\mu\nu}\rangle$ is fully simplified. The resulting energy density $\rho_{{\rm GW}}=\langle T_{00}\rangle $ of the GW is given by
\begin{align}
\rho_{\text{GW}} &\propto \int d \ln f \left[ 1 - \left( \frac{c I}{2 \pi f L} \right)^2 \right] f^{3} S_{h}(f), \label{T00intro} 
\end{align} 
where $f$ is the frequency of the massive KK modes but can be identified with the frequency of GW and $S_h(f)$ is the spectral density of the massive KK tensor modes. 
This result contains a characteristic factor
\begin{align}\label{fct}
C_I(f) \equiv  1- \left(\frac{Ic}{2\pi f L}\right)^2, 
\end{align}
 which determines the contributions of massive modes to the energy density of the GW only at the specific frequencies. From the positiveness of $C_I(f)$ in \eqref{fct}, we see that there exists a minimum value of the frequency in the GW detection. For instance, the minimum frequency is given by 
\begin{align}
f_{{\rm min}}\ge 4.8 \times 10^{11} (\text{Hz}),
\end{align}
for the upper bound on $L \le 10^{-4}{\rm m}$. The smaller the size of the extra dimension, the higher the minimum frequency. From the various stochastic background GWs observational limit~\cite{2003ApJ...599..806A,Shoda:2013oya,Coughlin:2014sca,Coughlin:2014xua,Aasi:2014zwg,Coughlin:2014boa,Adam:2014bub}, one can also estimate the amplitude of the massive KK modes that corresponds to this minimum frequency. We will present such order estimation in subsection \ref{subsec:32}. 

This paper is organized as follows. In section \ref{sec:DR}, we investigate the formalism of the effective 4-dimensional gravitational waves from the KK reduction of the 5-dimensional theory to the effective 4-dimensional gravity theory. The 4-dimensional massless graviton obtains the source from the massive modes. We show how to calculate this source term in the stochastic method in section \ref{sec:three}. We conclude in section \ref{sec:con}. 

\section{Dimensional Reduction with Massive KK Modes}
\label{sec:DR}

We consider the KK reduction of the 5-dimensional Einstein-Hilbert action without cosmological constant to obtain the 4-dimensional gravity theory. The KK reduction involves compactification of one spatial coordinate  on a circle of  radius $L$. For a finite $L$, the compactification results in 4-dimensional KK towers, which are the tensor modes $h^{I}_{\mu\nu}$,  the vector modes $v_\mu^I$, and scalar modes $s^I$ with $I = 0,1,2,\cdots$. The masses of these KK modes are proportional to $( \frac {I}{L})$.     
In the $L\to 0$ limit, all massive KK modes $(I\ne 0)$ are decoupled from the zero modes ($h_{\mu\nu}^0$, $v_\mu^0$, $s^0$). 
However, if $L$ is small but not zero, there exist  nontrivial couplings among the massless and few lower massive KK modes. 
In this paper, with an appropriate gauge choice, we get rid of the KK massive vector modes and investigate the couplings of the massless KK modes to the remaining massive KK modes. The result is the 4-dimensional linearized Einstein equation without the cosmological constant, but with the energy-momentum tensor determined by the massive KK modes. In this section, we present the detailed procedures.


In this paper, we consider the 5-dimensional Einstein-Hilbert action without cosmological constant \begin{equation} \label{EHact}
S = \frac{1}{16\pi G_5}\int d^4 x dy  
\sqrt{-\hat{g}} \,\, \hat{R}, 
\end{equation} 
where 
 $y$ denotes the coordinate of the extra dimension with  $S^1$ geometry. 
The 5-dimensional Einstein equation is
\begin{align}\label{eom1}
\hat R_{pq} - \frac12 \hat g_{pq}\hat R = 0.
\end{align}
In order to implement the KK reduction including massive KK modes, we introduce a metric fluctuation $\delta \hat g_{pq} = h_{pq}$ as, 
\begin{align}\label{hpq}
\hat g_{pq}=g_{pq}+h_{pq},
\end{align}
where $g_{pq}$ is the background metric. We plug \eqref{hpq} into \eqref{eom1} and expand the later up to quadratic order in the fluctuations. Since the background metric $g_{pq}$ is flat, the perturbed Einstein equation becomes
\begin{align}\label{lineq2}
&\delta \hat R_{pq}-\frac12\Big(g_{pq}g^{rs}\delta\hat R_{rs}+g^{rs}h_{pq}\delta\hat R_{rs}+g_{pq}\delta\hat g^{rs} \delta\hat R_{rs}\Big)
=0.
\end{align}
Inserting $\delta\hat g^{pq}=-h^{pq}+h^{pr}h_r^{~q}$ with $h^{pq} \equiv g^{pr}g^{qs}h_{rs}$ into \eqref{lineq2}, we obtain the equations for the fluctuations up to quadratic order in $h_{pq}$, 
\begin{align}\label{QdFl1}
&\nabla^r\nabla_ph_{rq}+\nabla^r\nabla_qh_{rp}-\nabla^2h_{pq}-\nabla_q\nabla_p h -g_{pq}\Big(\nabla^r\nabla^sh_{rs}-\nabla^2h\Big)
+Q_{pq}=0 
\end{align}
with
\begin{align}\label{QdTms}
Q_{pq}=&
-\nabla_r\Big(h^{rs}\big(\nabla_p h_{sq}
+\nabla_q h_{sp}-\nabla_s h_{pq}\big)\Big)+\frac12\nabla_qh^{rs}\nabla_p h_{rs}+h^{rs}\nabla_q\nabla_p h_{rs}
\\
&+\frac12\nabla^rh^s{}_s\big(\nabla_p h_{rq}+\nabla_q h_{rp}-\nabla_r h_{pq}\big)
+\nabla^r h^s{}_{q}\nabla_r h_{sp}-\nabla^r h^s{}_{q}\nabla_s h_{pr}
\nn\\
&+\frac12g_{pq}\nabla_r\Big(h^{rs}\big(2\nabla^t h_{st}-\nabla_s h^t{}_t\big)\Big)-\frac34g_{pq}\nabla^th^{rs}\nabla_t h_{rs}+\frac12g_{pq}\nabla^r h^{st}\nabla_s h_{tr}-\frac12g_{pq}h^{rs}\nabla^2 h_{rs}
\nn\\
&-\frac14g_{pq}\nabla^rh^s{}_s\big(2\nabla^s h_{rs}-\nabla_r h^t{}_t\big)
+\frac12g_{pq}h^{rs}\Big(\nabla^t\nabla_rh_{ts}
+\nabla^t\nabla_sh_{tr}-\nabla^2h_{rs}
-\nabla_r\nabla_sh^t{}_t\Big)
\nn \\
& + h_{pq} \nabla^2 h^r_{~r} - h_{pq} \nabla_r\nabla_s h^{rs},
\end{align}
where $Q_{pq}$ stands for  terms  that are quadratic in the fluctuations. Our goal in  this section is to implement the KK reduction and obtain the 4-dimensional linearized Einstein equation with a source from the 5-dimensional equation in \eqref{QdFl1}. The resulting equations contain information on the size of  extra dimension only if we take into account the massive KK modes.

The line element of the flat background metric with the compactified fifth direction has the form
\begin{align}\label{MinBac}
ds^2 = \hat g_{pq} dx^pdx^q =  \eta_{\mu\nu} dx^\mu dx^\nu +  dy^2,
\end{align} 
where $y=L\psi$ with  $0\le \psi < 2\pi$. The following procedure is identical to the Scherk-Schwarz formalism, which reduces $\hat{d}$-dimensional gravity to $d = \hat{d} -1$ dimensional effective theory\cite{Scherk_ea79}. Using the metric in \eqref{MinBac}, we split \eqref{QdFl1} into $(\mu,\nu)$, $(\mu,5)$, and $(5,5)$ components to obtain the equations of motion for the tensor, vector, and scalar fields. See Appendix \ref{QdEq}. Since the fifth-direction is a compactified, one can expand the fluctuations as 
\begin{align}\label{SHexp}
&h_{\mu\nu}(x,\psi)=h^{I}_{\mu\nu}(x)Y^{I}(\psi),\quad h^\mu{}_{\mu}(x,\psi) =h^{I}(x)Y^{I}(\psi),\quad \nn\\
&h_{\mu5}(x,\psi)=v_\mu^{0}(x)Y^{0}+v_\mu^{I}(x)\nabla_5Y^{I}(\psi),\quad h_{55}(x,\psi) =s^I(x)Y^{I}(\psi),
\end{align}
where $Y^I$'s with $I= 0,1,\cdots$ are the spherical harmonics on $S^1$, which satisfy the eigenequation $\nabla_5^2 Y^I=\frac{\partial^2}{\partial y^2} Y^I= \frac1{L^2}\frac{\partial^2}{\partial\psi^2} Y^I = \Lambda^IY^I~{\rm with}~\Lambda^I=-\frac {I^2}{L^2}$.   We normalized the spherical harmonics as,
\begin{align}\label{YIpsi}
Y^I(\psi) = \sqrt{2} \cos (I \psi)
\end{align} 
so that they obey the orthonormal condition
\begin{displaymath}
\frac1{2\pi}\int_0^{2\pi} d\psi\, Y^I(\psi) Y^J(\psi) =\left\{ \begin{array}{lr}
2  & (I=J = 0)\\
\delta^{IJ} & (I,J>0)
\end{array} \right..
\end{displaymath}

\subsection{Gauge fixing}\label{GIQ}
 Under an infinitesimal coordinate transformation $x'^p = x^p -\xi^p,$ the metric fluctuation transforms as 
\begin{align}\label{gautr1}
\delta_\xi h_{pq} = \nabla_p \xi_q + \nabla_q \xi_p.
\end{align}
Expanding the gauge function in terms of the spherical harmonics as 
\begin{align}\label{gautr2}
\xi_\mu(x,y) = \xi_\mu^I(x) Y^I(y), 
\qquad
\xi_5(x,y) = \xi^0(x) Y^0+\xi^I (x)\nabla_5 Y^I(y),
\end{align}
and using \eqref{SHexp} in \eqref{gautr1}, we obtain
\begin{align}\label{gatr5}
&\delta_\xi h_{\mu\nu}^I(x) = \nabla_\mu\xi^I_\nu (x)+ \nabla_\nu\xi^I_\mu (x),
\nn \\
&\delta_\xi v^0_\mu=\nabla_\mu \xi^0,\quad\delta_\xi v_\mu^{I}(x) = \nabla_\mu\xi^I(x) + \xi_\mu^I(x),~(I\neq 0) 
\nn \\
&\delta_\xi s^I(x)=  2\Lambda^{I}\xi^I(x).
\end{align}

Combining the tensor, the vector, and the scalar KK modes in \eqref{gatr5}, we obtain the gauge invariant massive KK tensor modes for $I\neq0$, 
\begin{align}\label{gainv0}
\hat h^I_{\mu\nu}(x) = h^{I}_{\mu\nu}(x) - \left(\nabla_\mu v_\nu^I(x) + \nabla_\nu v_\mu^I(x)\right) + \frac1{\Lambda^I}\nabla_\mu\nabla_\nu s^I(x).
\end{align}
In this paper, we fix the five gauge degrees of freedom for the non-zero modes as follows 
\begin{align}\label{gafix2}
v_\mu^I(x) = 0, \qquad s^I(x) = h^{I}(x), 
\end{align} 
where $I \neq 0$. 
Then the gauge invariant massive KK tensor modes are reduced to  
\begin{align}\label{gainv}
\hat h^I_{\mu\nu}(x) = h^{I}_{\mu\nu}(x)  + \frac1{\Lambda^I}\nabla_\mu\nabla_\nu h^I(x).
\end{align}
In the next subsection, we will see that the massive KK scalar modes $h^I$ and tensor modes $\hat h^I_{\mu\nu}$ play the role of the matter fields in the linearized Einstein equations of the massless 4-dimensional graviton mode.

In addition to the five gauge conditions in \eqref{gafix2}, we have five more constraints, which follow from the equations of motion for the non-zero modes. As a result, we get five physical degrees of freedom of the massive KK modes for every $I=1,2,\cdots$. See the discussion after \eqref{nz5}.

For the zero modes, we will use a different set of constraints to remove unphysical degrees of freedom. From the gauge transformations for the zero modes in \eqref{gatr5}, namely, $\delta_\xi h^0_{\mu\nu}=\nabla_\mu\xi^0_\nu+\nabla_\nu\xi^0_\mu$ and $\delta_\xi v^0_\mu=\nabla_\mu\xi^0$, we obtain five constraints. However, unlike the case of the non-zero modes, these five constraints are not suitable to  impose the conditions $v_\mu^0= 0$ and $s^0= h^0$.  Instead, we combine some of these gauge constraints with some of the five additional constraints, which are obtained from the equations of motion for the zero modes, to impose the transverse-traceless (TT) conditions on the 4-dimensional graviton mode. In the next subsection, we will introduce the 4-dimensional graviton mode $\hat h^0_{\mu\nu}$ and impose the TT gauge $\nabla^\mu\hat h^0_{\mu\nu}=0$, ~$g^{\mu\nu}\hat h^0_{\mu\nu}=0$.

\subsection{Linearized Einstein equations}
  
 In order to obtain the equations of motion for various KK modes, we split equation \eqref{QdFl1} into $(\mu,\nu)$, $(\mu,5)$,  $(5,5)$ components and use the expansions in terms of the spherical harmonics  given in \eqref{SHexp}. Then, the equations of motion for the $I^{th}$ KK modes are obtained by projecting on $Y^I$ or $\nabla^5 Y^I$.\footnote{See Appendix \ref{QdEq} for the details.}  In particular, the equations of motion for the zero modes ($I=0$) are given by 
 \begin{align}
&\nabla^\rho\nabla_\mu h^0_{\rho\nu}+\nabla^\rho\nabla_\nu h^0_{\rho\mu}-\square h^0_{\mu\nu}-\nabla_\mu\nabla_\nu(h^0+ s^0)
-g_{\mu\nu}\big(\nabla^\rho\nabla^\sigma h^0_{\rho\sigma}-\square(h^0+ s^0)\big)+Q^0_{\mu\nu}=0 \, , \label{QdEqmn-a}\\
&\nabla_\mu\nabla^\nu v^0_\nu-\square v^0_\mu+Q^0_\mu=0 \, ,\label{QdEqmn-e}\\
&\label{QdEqmn-d} 
\square h^0-\nabla^\rho\nabla^\sigma h^0_{\rho\sigma}+Q^0_{h}=0 \, ,
\end{align}
where $Q^0_{\mu\nu}$, $Q^0_\mu$, and $Q^0_h$ are the projections of the quadratic terms in \eqref{QdFl1} on $Y^0$ and they are composed of the massive KK modes $h^I_{\mu\nu}$, $h^I$ ($I\ne 0$). The explicit forms of $Q^0_{\mu\nu}$ and $Q^0_h$ are given in \eqref{Qhmn0} and \eqref{Qh0}, The explicit form of $Q^0_\mu$ is omitted because the equation of motion for the vector zero mode in \eqref{QdEqmn-e} is decoupled from the scalar and the tensor zero modes, which means $Q^0_\mu$ is  irrelevant to obtain the equation of motion for the 4-dimensional graviton. The equation of motion for the scalar zero mode $s^0$ is obtained by combining  the trace of \eqref{QdEqmn-a} with \eqref{QdEqmn-d}, which results in 
\begin{align}
&\square s^0 +Q^0_s=0,\label{QdEqmn-b}
\end{align}
where $Q_{s}^{0}$ is given in the equation \eqref{Qs0}.
Now we can plug \eqref{QdEqmn-d} and \eqref{QdEqmn-b} into \eqref{QdEqmn-a} to obtain  
\begin{align}\label{QdEqmn-c}
&\nabla^\rho\nabla_\mu h^0_{\rho\nu}+\nabla^\rho\nabla_\nu h^0_{\rho\mu}-\square h^0_{\mu\nu}-\nabla_\mu\nabla_\nu (h^0+s^0)-g_{\mu\nu}\big(Q^0_{h}+Q^0_s\big)+Q^0_{\mu\nu}=0.
\end{align}

Before we proceed further, we would like to comment on the fact that the massless KK zero modes cannot contribute to the quadratic terms in the above equations. Naively, the projection of \eqref{QdFl1} on $Y^0$ produces quadratic terms which are composed of both the massless and the massive KK modes.  For instance, the equation for the massless KK tensor mode can be rewritten as
\begin{align}
&\nabla^\rho\nabla_\mu h^0_{\rho\nu}+\nabla^\rho\nabla_\nu h^0_{\rho\mu}-\square h^0_{\mu\nu}-\nabla_\mu\nabla_\nu (h^0+s^0)=\widetilde Q^0_{\mu\nu}(h^I_{\mu\nu},h^I)+\widehat{Q}^0_{\mu\nu}(h^0_{\mu\nu},s^0,v_\mu^0)\, , \label{Boxh0exp}  
\end{align}    
where $\widetilde{Q}^{0}_{\mu\nu} (h_{\mu\nu}^{I}, h^{I})$ contains only the massive KK modes whereas $\widehat{Q}^{0}_{\mu\nu}(h^{0}_{\mu\nu} , s^{0},v^0_\mu)$ comes from the massless modes. Now if we introduce an order parameter of small fluctuations and write the zero mode as $h^{0}_{\mu\nu} = \lambda\,a^0_{\mu\nu} + {\cal O}(\lambda^2)$ then in order for $\widetilde{Q}^{0}_{\mu\nu} (h_{\mu\nu}^{I}, h^I)$ to be of the same order as the linear terms in \eqref{Boxh0exp}, the non-zero modes must be $h^{I}_{\mu\nu} = \sqrt{\lambda}\,a^I_{\mu\nu} + {\cal O}(\lambda)$. Thus, $\widehat{Q}^{0}_{\mu\nu}(h^{0}_{\mu\nu} , s^{0},v^0_\mu)$ will be second order in $\lambda$ and can be dropped from the linearized equations for zero modes. For this reason, the quadratic terms in the equations of motions for the massless KK modes can contain only the massive KK modes. 

We can simplify the quadratic terms in the above equations by using the equations of motion for the massive modes. As those quadratic terms are already second order in the fluctuations, the linear order of the massive modes is sufficient for the purpose of such simplification. After fixing the gauge as in \eqref{gafix2},  the linear order equations of motion for the massive modes, which are obtained by projecting \eqref{QdFl1} on $Y^I$ or $\nabla^5 Y^I$, are given by 
\begin{align}
&\nabla_\mu\nabla^\rho h^I_{\rho\nu}+\nabla_\nu \nabla^\rho h^I_{\rho\mu}-(\square+\Lambda^I)h^I_{\mu\nu}-2\nabla_\mu\nabla_\nu h^I-g_{\mu\nu}\Big[\nabla^\rho\nabla^\sigma h^I_{\rho\sigma}-\Lambda^I h^I-2\square h^I\Big]=0 \, , \label{nz1} \\
&\square h^I=-\Lambda^I h^I, \label{nz2}\\
&\nabla^\rho h^I_{\rho\mu}=\nabla_\mu h^I.\label{nz4}
\end{align}
Combining \eqref{nz2} and  \eqref{nz4} we obtain one more constraint, which is
\begin{align}\label{nz5}
h^I=-\frac1{\Lambda^I}\nabla^\mu\nabla^\nu h^I_{\mu\nu}.
\end{align}
The five constraints from \eqref{nz4} and  \eqref{nz5} will eliminate five degrees of freedom among the fifteen degrees of freedom in $h^I_{pq}$. Additional five degrees of freedom are removed by the gauge fixing in \eqref{gafix2}. This will leave us with five physical degrees of freedom for the non-zero modes at each $I=1,2,\cdots$.

For the zero modes, the TT gauge, which was introduced in the previous subsection, removes five unphysical degrees of freedom from $\hat h^0_{\mu\nu}$. We then use the remaining five constraints to remove three more unphysical degrees of freedom from $\hat h^0_{\mu\nu}$ and two unphysical degrees of freedom from $v^0_\mu$.    As a result, we have five physical degrees of freedom for the zero modes as well.  For clarity, we summarize the physical degrees of freedom for the massless and massive modes in  table 1.    
\begin{table}[!h]
\begin{center}
\begin{tabular}{|c|c|c|c|c|c|}
\hline
\multirow{3}{*}{}            & & & &  &  \\ 
Modes          & 5-dim & d.o.f  & 4-dim & d.o.f & mass   \\ 
&  fields &  & fields &   &    \\ \hline
\multirow{3}{*}{} &   &  & $\hat h_{\mu\nu}^{0}$&  2 &   0  \\ \cline{4-6} 
$I=0$  & $h_{pq}^{0}$  & 5 & $v_{\mu}^{0}$  & 2  & 0  \\ \cline{4-6} 
                  &   & & $s^{0}$ & 1 & 0  \\ \hline
\multirow{2}{*}{} $I\neq 0$ & $h_{pq}^{I}$  & 5 &  ${\hat h}_{\mu\nu}^{I}$ & 4 & $I/L$ \\ \cline{4-6} 
     &   &  & $h^{I}$ & 1  & $I/L$ \\ \hline
\end{tabular}
\caption{The physical degrees of freedom of five-dimensional fields and four-dimensional fields. Instead of two physical d.o.f of $h_{\mu\nu}$, we choose four physical d.o.f for $\hat{h}_{\mu\nu}^{I}$ by removing two d.o.f of $v_{\mu}^{I}$.}
\label{tab1}
\end{center}
\end{table}

Now inserting \eqref{nz2} and \eqref{nz4} into \eqref{nz1}, we obtain the linearised equation for the massive KK tensor mode
\begin{align}
\square h^I_{\mu\nu}=-\Lambda^Ih^I_{\mu\nu} \, . \label{nz3}
\end{align}
The mass of these KK modes is inversely proportional to the size of the compactified fifth dimension. Thus, we can constrain this size if we can measure the mass of the KK modes. We will discuss this in section \ref{sec:three}.

Finally, we can use \eqref{nz2}, \eqref{nz4} and  \eqref{nz3} to simplify the quadratic terms in  \eqref{QdEqmn-b} and \eqref{QdEqmn-c}. After the simplification, we combine \eqref{QdEqmn-b} and \eqref{QdEqmn-c} to obtain the equation of motion for the 4-dimensional massless graviton mode,
\begin{align}\label{QdEqhmn2}
&L_E\hat h^0_{\mu\nu}-\frac14\nabla^\sigma h^I_{\rho\nu}\nabla^\rho h^I_{\sigma\mu}+\frac14\nabla^\sigma h^{I\rho}_{\nu}\nabla_\sigma h^I_{\mu\rho}+\frac18\nabla_\mu h^{I\rho\sigma}\nabla_\nu h^I_{\rho\sigma}+\frac14 h^{I\rho\sigma}\nabla_\nu \nabla_\mu h^I_{\rho\sigma}+\frac14h^{I\rho\sigma}\nabla_\rho \nabla_\sigma h^I_{\mu\nu}\nn\\
&-\frac14h^{I\rho\sigma}\big[\nabla_\rho\nabla_\mu h^I_{\sigma\nu}+\nabla_\rho\nabla_\nu h^I_{\sigma\mu}\big]-\frac{\Lambda^I}4 h^{I\rho}_{\nu}h^I_{\mu\rho}+\frac14\Lambda^Ih^I h^I_{\mu\nu}+\frac1{8}\nabla_\mu h^I\nabla_\nu {h^I}+\frac1{4} h^I \nabla_\mu\nabla_\nu h^I\nn\\
&+g_{\mu\nu} \Big[\frac{\Lambda^I}{16} h^{I\rho\sigma}h^I_{\rho\sigma}+\frac18  h^{I\rho\sigma}\nabla_\rho\nabla_\sigma h^I+\frac1{8} \nabla^\rho h^I\nabla_\rho h^I- \frac{\Lambda^I}{16} h^I h^I\Big]=0,
\end{align}
where we have defined the 4-dimensional massless graviton $\hat h^0_{\mu\nu}=h^0_{\mu\nu}+\frac12 g_{\mu\nu}s^0$. Here we introduced the Einstein operator
\begin{align}
L_E A_{\mu\nu}=\frac12\Big(-\square A_{\mu\nu}+\nabla^\rho\nabla_\mu A_{\rho\nu}+\nabla^\rho\nabla_\nu A_{\rho\mu}-\nabla_{\mu}\nabla_{\nu}A^\rho_\rho\Big).
\end{align}
The result in \eqref{QdEqhmn2} is the linearised Einstein equation with the source for the 4-dimensional graviton. One can see that the massive KK modes become the source of the 4-dimensional massless graviton. Compared to the usual 4-dimensional graviton which does not contain the source for the linear order perturbation, the effective 4-dimensional massless graviton obtains the source from the massive KK modes in the compactified 5-dimensional Universe. 

We can write \eqref{QdEqhmn2} more formally as
\begin{align}\label{LinEq0}
L_E \hat h^0_{\mu\nu} = 8\pi G_N\left(T_{\mu\nu}[h^{I}_{\rho\sigma}, h^{I}] -\frac12g_{\mu\nu} T [h^{I}_{\rho\sigma}, h^{I}] \right) \, , 
\end{align} 
where the energy-momentum tensor is given by
\begin{align}\label{Tmn}
T_{\mu\nu}[h^{I}_{\rho\sigma}, h^{I}] = -\frac1{32\pi G_N}\bigg\{&\nabla^\sigma h^{I\rho}_{\nu}\nabla_\sigma h^I_{\mu\rho}-\nabla^\sigma h^I_{\rho\nu}\nabla^\rho h^I_{\sigma\mu}+\frac12\nabla_\mu h^{I\rho\sigma}\nabla_\nu h^I_{\rho\sigma}\nn\\
&+ h^{I\rho\sigma}\nabla_\nu \nabla_\mu h^I_{\rho\sigma}+h^{I\rho\sigma}\nabla_\rho \nabla_\sigma h^I_{\mu\nu}-h^{I\rho\sigma}\big[\nabla_\rho\nabla_\mu h^I_{\sigma\nu}+\nabla_\rho\nabla_\nu h^I_{\sigma\mu}\big]\nn\\
&-\Lambda^I h^{I\rho}_{\nu}h^I_{\mu\rho}+\Lambda^Ih^I h^I_{\mu\nu}+\frac1{2}\nabla_\mu h^I\nabla_\nu {h^I}+ h^I \nabla_\mu\nabla_\nu h^I\nn\\
-\frac{g_{\mu\nu}}{4} \Big[3 \nabla^\tau h^{I\rho\sigma}\nabla_\tau h^I_{\rho\sigma}&-2 \nabla^\rho h^{I\sigma\tau}\nabla_\sigma h^I_{\rho\tau}-3\Lambda^I h^{I\rho\sigma}h^I_{\rho\sigma}+3 \nabla^\rho h^I\nabla_\rho h^I-\Lambda^I h^I h^I\Big]\bigg\}.
\end{align} 
We note that even though $h^I$ appears in the energy-momentum tensor, it is not an independent degree of freedom since it is expressed in terms of $h^I_{\mu\nu}$ as in \eqref{nz5}.

 For later convenience, we rewrite \eqref{Tmn} in terms of the gauge invariant massive KK tensor mode defined in \eqref{gainv}. From \eqref{nz2}, we obtain
$\nabla^\mu\hat h^I_{\mu\nu}=0$ and $g^{\mu\nu}\hat h^I_{\mu\nu}=0$. Thus,  
\begin{align}\label{Tmunu}
T_{\mu\nu}[\hat{h}^{I}_{\rho\sigma}, h^{I}] = &-\frac1{16\pi G_N}\bigg\{\frac12\hat h^{I\rho\sigma}\nabla_\rho \nabla_\sigma \hat h^I_{\mu\nu}-\frac12\hat h^{I\rho\sigma}\nabla_\rho\nabla_\mu \hat h^I_{\sigma\nu}-\frac12\hat h^{I\rho\sigma}\nabla_\rho\nabla_\nu \hat h^I_{\sigma\mu}+\frac12 \hat h^{I\rho\sigma}\nabla_\nu \nabla_\mu \hat h^I_{\rho\sigma}
\nn \\
&+\frac12\nabla^\sigma \hat h^{I\rho}_{\nu}\nabla_\sigma \hat h^I_{\mu\rho}-\frac12\nabla^\sigma \hat h^I_{\rho\nu}\nabla^\rho \hat h^I_{\sigma\mu}+\frac14\nabla_\mu \hat h^{I\rho\sigma}\nabla_\nu \hat h^I_{\rho\sigma}-\frac{\Lambda^I}2 \hat h^{I\rho}_{\nu}\hat h^I_{\mu\rho} 
\nn\\
&-\frac1{4\Lambda^I}\nabla_\mu\nabla^\rho\nabla^\sigma h^I\nabla_\nu \hat h^I_{\rho\sigma}-\frac1{4\Lambda^I}\nabla_\nu\nabla^\rho\nabla^\sigma h^I\nabla_\mu \hat h^I_{\rho\sigma}-\frac1{2\Lambda^I}\nabla^\rho\nabla^\sigma h^I\nabla_\rho\nabla_\sigma \hat h^I_{\mu\nu}\nn\\
&+\frac1{2\Lambda^I}\nabla^\rho\nabla^\sigma h^I\nabla_\rho\nabla_\mu \hat h^I_{\sigma\nu} + \frac1{2\Lambda^I}\nabla^\rho\nabla^\sigma h^I\nabla_\rho\nabla_\nu \hat h^I_{\mu\sigma}-\frac1{2\Lambda^I}\nabla^\rho\nabla^\sigma h^I\nabla_\mu\nabla_\nu \hat h^I_{\rho\sigma}
\nn \\
&+\frac1{2\Lambda^I}\Lambda^I\nabla_\mu\nabla^\rho h^I\hat h^I_{\rho\nu}+\frac1{2\Lambda^I}\Lambda^I\nabla_\nu\nabla^\rho h^I\hat h^I_{\mu\rho}+\frac12\Lambda^I h^I \hat h^I_{\mu\nu}
\nn\\
&+\frac1{4(\Lambda^I)^2}\nabla_\mu \nabla^\rho\nabla^\sigma h^I\nabla_\nu \nabla_\rho\nabla_\sigma h^I -\frac1{2\Lambda^I}\nabla_\mu\nabla^\rho h^I\nabla_\nu\nabla_\rho h^I+\frac1{4}\nabla_\mu h^I\nabla_\nu {h^I}
\nn\\
&-\frac12g_{\mu\nu}\Big[\frac34\nabla_\tau \hat h^{I\rho\sigma}\nabla^\tau \hat h^I_{\rho\sigma}-\frac12\nabla^\sigma \hat h^I_{\rho\tau}\nabla^\rho \hat h^{I\tau}_{\sigma}-\frac{3\Lambda^I}4\hat h^{I\rho\sigma}\hat h^I_{\rho\sigma} -\frac1{2\Lambda^I}\nabla^\tau\nabla^\rho\nabla^\sigma h^I\nabla_\tau \hat h^I_{\rho\sigma}
\nn \\
&+\frac3{2}\nabla^\rho\nabla^\sigma h^I \hat h^I_{\rho\sigma}+\frac1{4(\Lambda^I)^2}\nabla^\tau \nabla^\rho\nabla^\sigma h^I\nabla_\tau \nabla_\rho\nabla_\sigma h^I-\frac3{4\Lambda^I}\nabla^\rho\nabla^\sigma h^I\nabla_\sigma\nabla_\rho h^I
\nn\\
&+\frac3{4}\nabla_\rho h^I\nabla^\rho {h^I} -\frac{\Lambda^I}4 h^I h^I\Big]\bigg\}.
\end{align}
Even though it is redundant to rewrite the $T_{\mu\nu}$ in \eqref{Tmunu}, this form will further simplify our calculations in the subsequent sections. For comparison, we shortly review how to obtain the effective energy-momentum tensor of the four-dimensional gravitational waves on the flat universe in Appendix \ref{sec:AppB}. It is important to notice that the effective energy-momentum tensor obtained there is  different from the above energy-momentum tensor of our model.

\section{Effect of the Extra Dimension in the GW detection}
\label{sec:three}

In this section, we show the effect of massive KK modes on the dynamics of the four-dimensional graviton.  
  In order to compare with the observation, we derive the analytic form of the stochastic average of the energy-momentum tensor obtained in the previous section. We perform the order estimation to put a new constraint on the size of the extra dimension. It is expected that there will be more accurate GW data available in the future and we believe our results provide a tangible formalism on how to use  the GW data to get some information about the extra dimension.

\subsection{Stochastic average of the energy-momentum tensor}
\label{subsec:31}
The energy in GW is described by averaging the energy-momentum tensor over several wavelengths or periods. Thus, we need to average the energy-momentum tensor in our model given in (\ref{Tmunu}), which is built from the non-zero modes, $h^I$ and $\hat{h}^{I}_{\mu\nu}$. 
Inserting the plane wave solution  $\hat{h}_{\mu\nu}^{I} = {A}_{\mu\nu}^I e^{i p_{\mu} x^{\mu}}$ into (\ref{nz3}), we obtain
\begin{align} \Box \hat{h}_{\mu\nu}^I + \Lambda^{I} \hat{h}_{\mu\nu}^I &= \left( - p_{\mu} p^{\mu} + \Lambda^{I} \right) \hat{h}_{\mu\nu}^I = 0 \,\,\, \longrightarrow \,\,
m_{\text{I}}^2 = -\frac{\Lambda^{I} \hbar^2}{c^2} = \left( \frac{I \hbar}{c L} \right)^2 \,\, , \label{ms} \end{align}
where we have used the dispersion relation 
\begin{align}\label{dis-rel}
\frac{\omega_I^2}{c^2} - |\vec{k}_I|^2 = \frac{m_I^2 c^2}{\hbar^2}
\end{align}
 and the definition $\Lambda^{I} = - \frac{I^2}{L^2}$.  We can regard both $h^I$ and $\hat{h}^{I}_{\mu\nu}$ as random variables, which can be identified with a stochastic background. Thus, using the dispersion relations, we calculate the ensemble average of the energy-momentum tensor. In order to perform the time average, we use the plane wave expansion of $h^I(\vec{x},t)$ in terms of the Fourier transformation \cite{Maggiore:1999vm,Maggiore2008}. 
\begin{align} 
h^I({\vec{x}},t) & = \int\frac{d^3 k}{(2\pi)^3} \left[ \tilde{B}^{I} ( \vec{k}) e^{i k_\mu x^\mu} + \tilde{B}^{I\ast} ( \vec{k}) e^{-i k_\mu x^\mu} \right], \label{sx3D} 
\end{align}
where the four-vectors in the exponent are $k^\mu=(\omega/c,\vec{k}\,),~x^\mu=(ct,\vec{x}\,)$ and we have suppressed the index $I$ in ($\omega_I,~\vec{k}_I$)  for simplicity. 
Using the dispersion relation in \eqref{dis-rel} we can write $d^3 k$ as
\begin{align}
d^3 k &= |\vec{k}|^2 d |\vec{k}| d^{2} \hat{n} = \left( 1 + \frac{c^2}{\omega^2} \Lambda^{I} \right)^{\frac{1}{2}} \frac{\omega^{2}}{c^{3}} d \omega d^{2} \hat{n} = \sqrt{1- \left( \frac{c I}{2\pi f L} \right)^2} \left(\frac{2\pi}{c} \right)^3 f^2 df d^{2} \hat{n}, \label{d3k} 
\end{align}
where $d^2\hat n=-d(cos\theta)d\phi$ denotes the integration over the solid angle and in the last step we have replaced $\omega$ by $2\pi f$.
Similarly
\begin{align}
k_\mu x^\mu=-\omega t+|\vec{k}|\hat n \cdot \vec{x}=-2\pi f\left(t-\sqrt{1- \left( \frac{c I}{2\pi f L} \right)^2}\frac{\hat n \cdot \vec{x}}c\right).\label{Exp} 
\end{align}
Using \eqref{d3k} and \eqref{Exp},  \eqref{sx3D} becomes
\begin{align}
&h^I(\vec{x},t)= \int_{\epsilon}^{\infty} df \int d^{2} \hat{n} \sqrt{1 - \left(\frac{Ic}{2\pi f L} \right)^2} \left[\frac{f^2}{c^3}\tilde{B}^{I} ( f, \hat{n}) e^{-2\pi i f\left(t-\sqrt{1- \left( \frac{c I}{2\pi f L} \right)^2}\frac{\hat n \cdot \vec{x}}c\right)} + {\rm (c.c)} \right] \nonumber \\ 
                 &\equiv  \int_{-\infty}^{\infty} df \int d^2 \hat{n} \sqrt{1 - \left(\frac{Ic}{2\pi f L} \right)^2}  \tilde{h}^{I} ( f, \hat{n}) e^{-2\pi i f\left(t-\sqrt{1- \left( \frac{c I}{2\pi f L} \right)^2}\frac{\hat n \cdot \vec{x}}c\right)}, \label{hIint}  
 \end{align}
where $\epsilon = mc^2/(2\pi \hbar)$ and in the second line we used $\tilde{B}^{I\ast}(f,\hat n) = \tilde{B}^{I}(-f,\hat n)$, (c.c) means the complex conjugate, and also introduced  $\tilde{h}^{I}(f, \hat{n})\equiv \frac{f^2}{c^3} \tilde{B}^{I}(f, \hat{n})$. We emphasize that the integral range in \eqref{hIint} is not well defined for $|f| < \epsilon$ because it will produce tachyonic modes as shown in \eqref{dis-rel}. The real integral ranges in the second equality of \eqref{hIint} are $[-\infty, -\epsilon]$ and $[\epsilon, \infty]$. However, this detail is not important because as we will see later, the quantities we need to compare with the data are the integrands of \eqref{hIave}, \eqref{rhodef}, and \eqref{Omegagwdef}. 

From the above plane wave solution and assumption on the stochastic properties of the non-zero modes, we can replace the ensemble average of the energy-momentum tensor with the temporal average. This is given by
\begin{align}          
&\Bigl \langle h^{I\ast}({\vec x},t) h^I({\vec x},t) \Bigr \rangle =  \int_{-\infty}^{\infty} df \int d^{2} \hat{n} \sqrt{1 - \left(\frac{Ic}{2\pi f L} \right)^2}  \int_{-\infty}^{\infty} df' \int d^{2} \hat{n}' \sqrt{1 - \left(\frac{Ic}{2\pi f' L} \right)^2} \nonumber \\
 &\times \left \langle \tilde{h}^{I\ast}(f,\hat{n}) \tilde{h}^{I}(f',\hat{n}')\right\rangle e^{2\pi i f\left(t-\sqrt{1- \left( \frac{c I}{2\pi f L} \right)^2}\frac{\hat n \cdot \vec{x}}c\right)}e^{-2\pi i f'\left(t-\sqrt{1- \left( \frac{c I}{2\pi f' L} \right)^2}\frac{\hat n' \cdot \vec{x}}c\right)}. \label{ssave} 
\end{align}
Note that since $h^I({\vec x},t)$ is a real mode, we could write $\Bigl \langle h^{I\ast}({\vec x},t) h^I({\vec x},t) \Bigr \rangle$ instead of $\Bigl \langle h^{I}({\vec x},t) h^I({\vec x},t) \Bigr \rangle$, because the former way of writing is commonly used in literature. 
Now we use the definition of the stochastic property $\left \langle \tilde{h}^\ast(f,\hat{n}) \tilde{h}(f',\hat{n}') \right \rangle \equiv \frac{1}{2}  \delta(f-f') \frac{\delta^{2}(\hat{n}, \hat{n}')}{4\pi} S_{h}(f)$. The function $S_{h}(f)$ is called the spectral density of the stochastic background $h^I({\vec x},t)$.  Then we obtain
\begin{align}          
\Bigl \langle h^{I\ast}({\vec x},t) h^I({\vec x},t) \Bigr \rangle &=  \frac12\int_{-\infty}^{\infty} df \int d^{2} \hat{n} \left[1 - \left(\frac{Ic}{2\pi f L} \right)^2\right]\frac{S_h(f)}{4\pi}\nn\\
& =\int_{\epsilon}^{\infty} df \left[1 - \left(\frac{Ic}{2\pi f L} \right)^2\right]S_h(f),\label{hIave} 
\end{align}
where we have used $S_h(-f)=S_h(f)$ in the second line.

One can repeat the same process for the $\hat{h}_{\rho\sigma}^{I}$ to obtain 
\begin{align}
\hat{h}_{\rho\sigma}^{I} ({\vec x},t) 
&= \int \frac{d^3 k}{(2\pi)^3} \left[ {\cal A}^{I}_{\rho\sigma}(\vec{k}) e^{i k_\mu x^\mu} + {\cal A}_{\rho\sigma}^{I\ast}(\vec{k}) e^{-i k_\mu x^\mu} \right] \nonumber  \\ 
&=  \int_{-\infty}^{\infty} df \int d^2 \hat{n} \sqrt{1 - \left(\frac{Ic}{2\pi f L} \right)^2}  \tilde{h}_{\rho\sigma}^{I} ( f, \hat{n}) e^{-2\pi i f\left(t-\sqrt{1- \left( \frac{c I}{2\pi f L} \right)^2}\frac{\hat n \cdot \vec{x}}c\right)},    \label{hFT32}
\end{align}
where we have define $\tilde{h}_{\rho\sigma}^{I}(f, \hat{n}) \equiv\frac{f^2}{c^3} {\cal A}_{\rho\sigma}(f, \hat{n})$. If we adopt the stochastic background of the GW generated from $h_{\mu\nu}^{I}$ then 
\begin{align}
\tilde{h}_{\mu\nu}^{I}(f,\hat{n}) &= \sum_{A} \tilde{h}_{A}^{I}(f,\hat{n}) \epsilon_{\mu\nu}^{A}, \label{tildehA}
\end{align}
where $\epsilon_{\mu\nu}^{A}$'s are the polarization tensors and $A$ labels the polarization modes ($A = 1, 2, \cdots, 4$) as shown in Table 1. If one assumes that the stochastic backgrounds of GW are stationary, then the ergodic average becomes
\begin{align}
\left\langle \tilde{h}^{I\ast}_{\rho\sigma}(f, \hat{n}) \tilde{h}^{\rho\sigma I } (f', \hat{n}') \right \rangle &\equiv \sum_{A} \epsilon_{\rho\sigma}^{A}  \sum_{A'} \epsilon^{A' \rho\sigma} \left\langle \tilde{h}_{A}^{I\ast}(f, \hat{n}) \tilde{h}_{A'}^{I} (f', \hat{n}') \right \rangle  \nn\\
&=\sum_{A} \epsilon_{\rho\sigma}^{A}  \epsilon^{A \rho\sigma} \delta(f - f') \frac{\delta^{2}(\hat{n},\hat{n}')}{4\pi} \frac{1}{2} S_{h^{I}}(f),  \label{SD}
\end{align}
where we have used $\left\langle \tilde{h}_{A}^{I\ast}(f, \hat{n}) \tilde{h}_{A'}^{I} (f', \hat{n}') \right \rangle \equiv \delta(f - f') \frac{\delta^{2}(\hat{n},\hat{n}')}{4\pi} \delta_{AA'} \frac{1}{2} S_{h^{I}}(f)$. 
Now using \eqref{hFT32} and \eqref{SD}, the average for the tensor modes becomes
\begin{align}
\Bigl \langle \hat h_{\rho\sigma} (\vec x, t) \hat h^{\rho \sigma} (\vec x,t) \Bigr \rangle 
&= \sum_{A} \epsilon_{\rho\sigma}^{A}  \epsilon^{A \rho\sigma} \int_{\epsilon}^{\infty} df \left[ 1 - \left( \frac{c I}{2 \pi f L} \right)^2 \right]  S_{h^{I}}(f) \nonumber \\
&= \sum_{A} \epsilon_{\rho\sigma}^{A}  \epsilon^{A \rho\sigma} \int_{f=\epsilon}^{\infty} d \ln f \left[ 1 - \left( \frac{c I}{2 \pi f L} \right)^2 \right] f S_{h^{I}}(f) \nonumber  \\
&\equiv \frac{\sum_{A} \epsilon_{\rho\sigma}^{A} \epsilon^{A \rho\sigma}}{2}   \int_{f=\epsilon}^{\infty} (d\ln f)h_{c}^{2}(f) \, , \label{hhave2}
\end{align}
where the definition of characteristic strain is $h_{c} = \sqrt{2 (1 - (c I / 2 \pi f L)^{2}) f S_{h^{I}}}$. For $I=0$, we recover the usual 4-dimensional characteristic strain $\sqrt{2 f S_{h^{I}}}$. The dimension of $S_{h^{I}}$ is [Hz$^{-1}$] and $h_{c}$ is dimensionless. 

Finally, we can calculate $\left < T_{\mu\nu} \right>$ using the above stochastic properties of $h^I$ and $\hat{h}^{I}_{\mu\nu}$.   The energy-momentum tensor for our model is given in \eqref{Tmunu} and it looks very complicated. However, once we substitute the above plane wave solutions for the massive scalar and tensor modes, most of the terms are either vanishing because of the transverse condition $\nabla^\mu\hat h^I_{\mu\nu}\to k^\mu\hat h^I_{\mu\nu}=0$ or they cancel each other because any two contracted covariant derivatives of the form $\big\langle\nabla^\rho (\nabla\cdots h^I)\nabla_\rho(\nabla\cdots h_{\mu\nu}^I)\big\rangle$ becomes $\Lambda^I \big\langle(\nabla\cdots h^I)(\nabla\cdots h_{\mu\nu}^I)\big\rangle$, where the ellipses denote multiple covariant derivatives.   As a result, we get a very remarkable simplification and only two terms in \eqref{Tmunu} survive,
\begin{align}
\left \langle T_{\mu\nu} \right \rangle &= -\frac{c^4}{16 \pi G} \left \langle \frac{1}{2} \hat h^{I \rho \sigma}(x^{\alpha}) \nabla_{\mu} \nabla_{\nu} \hat h^{I}_{\rho\sigma} (x^{\alpha}) + \frac{1}{4} \nabla_{\mu} \hat h^{I \rho \sigma}(x^{\alpha}) \nabla_{\nu} \hat h^{I}_{\rho\sigma} (x^{\alpha})\right \rangle \nonumber \\ 
&= \frac{c^4}{16 \pi G} \frac{\sum_{A} \epsilon_{\rho\sigma}^{A} \epsilon^{A \rho\sigma}}{4} \int_{f=\epsilon}^{\infty} d \ln f\, k_{\mu}k_{\nu} \left[ 1 - \left( \frac{c I}{2 \pi f L} \right)^2 \right] f S_{h^{I}}(f) \label{TmunuSto} \, .
\end{align}
Thus, the energy density of the GW $\big(\rho_{\text{gw}}= \langle T_{00}\rangle\big)$ is given by
\begin{align}
\rho_{\text{gw}} = \int_{f=\epsilon}^{\infty} d \ln f \frac{d \rho_{\text{gw}}}{d \ln f}= \frac{c^2 \pi}{4 G} \frac{\sum_{A} \epsilon_{\rho\sigma}^{A} \epsilon^{A \rho\sigma}}{4} \int_{f=\epsilon}^{\infty} d \ln f \left[ 1 - \left( \frac{c I}{2 \pi f L} \right)^2 \right] f^{3} S_{h^{I}}(f)  \label{rhodef}, 
\end{align}
where the first step amounts to expressing the energy density of GW as an integral over $d \ln f$ of some spectral density and the second step is substituting $\langle T_{00}\rangle$ from \eqref{TmunuSto}. One can also define the logarithmic derivative of the energy density contrast of the GW as
\begin{align}
\Omega_{\text{gw}}(f) &\equiv \frac{1}{\rho_{\text{cr}}} \frac{d \rho_{\text{gw}}}{d \ln f} \label{Omegagwdef} =\frac{2 \pi^2}{3 H_0^2} \frac{\sum_{A} \epsilon_{\rho\sigma}^{A} \epsilon^{A \rho\sigma}}{4} \left[ 1 - \left( \frac{c I}{2 \pi f L} \right)^2 \right] f^{3} S_{h^{I}}(f) \nn\\
&= \frac{2 \pi^2}{3 H_0^2} f^{2} h_{c}(f)^{2} = \frac{8 \pi^2}{3 H_0^2} f^{4} |\tilde{h}(f)|^{2}  \, ,
\end{align}
where $\rho_{\rm cr}\equiv \frac{3c^2}{8\pi G} H_0^2$ is the critical energy density, $H_0$ is the present value of Hubble constant, and $|\tilde{h}(f)| = h_{c}(f)/(2 f)$ is the so-called frequency-domain strain. We have also used the fact that $\sum_{A} \epsilon_{\rho\sigma}^{A} \epsilon^{A \rho\sigma} = 8$ in the second equality because the four massive tensor modes $\hat{h}_{\mu\nu}^{I}$ contribute to the sources of the stochastic background GW.

\subsection{Order estimations}
\label{subsec:32}

In this section, we elaborate on the observational applications of our model by using both GW experiments and particle physics ones.

\subsubsection{Gravitational wave applicataion}
\label{subsubsec:321}

One of the main properties of our model is the existence of the lower limit of the frequency which can contribute as a source of the 4-dimensional GWs. From \eqref{Omegagwdef} we notice that we first need to make sure $1 - \frac{I c}{2 \pi f L}$ is positive. This means that the KK modes below a specific frequency cannot contribute as the source of the effective 4-dimensional massless gravitational waves. Thus, we obtain the lower limit on the frequency $f_{\text{min}}$ for the first massive mode ($I=1$) for a given upper bound on the size of the extra dimension $L_{\text{max}}$, 
\begin{align}
f_{\text{min}} \geq \frac{ c}{2 \pi L_{\text{max}}}. \label{fl} 
\end{align}
There have been various upper limits on the size of the extra dimension depending on the methods of measurements \cite{2018PhRvD..98c0001T} . One way to put the limit on the $L_{\text{max}}$ is the deviations of the Newtonian gravitational law and it gives the value as around $100 \mu$m~\cite{2018PhRvD..98c0001T}. In this case, we can put the lower limit on the frequency as ${\cal O}\big(10^{11}\big)$ Hz. Another method is based on the Randal-Sundrum (RS) model  which proposes the $1/$TeV ({\it i.e.} $10^{-18}$m) scale warped extra dimension in order to explain the hierarchy of the electroweak scale \cite{Randall:1999ee}. Thus, the RS model estimates the lowest value of the frequency to be ${\cal O}\big(10^{25}\big)$ Hz.
The smaller the size of the extra dimension, the higher the frequency of the modes that can contribute as the source of the massless 4-dimensional GWs. Therefore, we propose high-frequency detectors in order to measure the very small extra dimension by using the GWs.

Next, we use the observational limit on the magnitude of the dimensionless energy density in GWs, $\Omega_{\text{gw}}(f)$, to constrain the magnitude of characteristic strain $h_{c}$ by using \eqref{Omegagwdef}. 
There are various limits on the $\Omega_{\text{gw}}(f)$ at fixed frequencies from various observations \cite{2003ApJ...599..806A,Shoda:2013oya,Coughlin:2014sca,Coughlin:2014xua,Aasi:2014zwg,Coughlin:2014boa,Adam:2014bub}. We show these in Table \ref{tab2}. If we adopt the Planck result \cite{Adam:2014bub} and use the fact that $\frac{\Omega_{\rm gw}(f_{\rm Pl})}{ f_{\rm Pl}^{2}} = \frac{\Omega_{\rm gw}(f_{\rm KK})}{f_{\rm KK}^{2}}$,  then we can estimate the magnitude of $\tilde{h}$ as
\begin{align}
\tilde{h} = \frac{h_{c}}{2f} = \sqrt{\frac{3 \Omega_{\text{gw}} }{8}} \frac{H_0}{\pi f^2} = \frac{h_{c}({\rm Pl})}{2 f({\rm KK})} \leq \frac{3.78 \times10^{-25}}{2 \times 10^{11}} \, \simeq 10^{-36} \, (\text{Hz}^{-1}) \, . \label{hcexp}
\end{align} 

\begin{table}[h!]
\centering
\resizebox{\textwidth}{!}{%
\begin{tabular}{|c|c|c|c|c|c|}
\hline
$\Omega_{\text{gw}}(f)$ & $f$(Hz) &  Experiments     & $h_{c}$             & $\tilde{h}$            & Ref                       \\
\hline
0.044                              & $3 \times 10^{-4}$  & Cassini & $2 \times 10^{-15}$ & $8.32 \times 10^{-11}$ & \cite{2003ApJ...599..806A} \\
$3.88 \times 10^{17}$ & $0.035 - 0.830$        & TOBA   &  $7.71 \times 10^{-9}$ & $7.71 \times 10^{-8}$ & \cite{Shoda:2013oya}   \\
$1.2 \times 10^{8}$ & $0.05 - 0.1$        & Seismic   &  $1.29 \times 10^{-13}$ & $8.61 \times 10^{-13}$ & \cite{Coughlin:2014sca}  \\
$0.035-0.15$ & $0.005 - 0.3$        & Earth's ring   &  $5.51 \times 10^{-19}$ & $9.19 \times 10^{-19}$ & \cite{Coughlin:2014xua}  \\
$5.6\times10^{-6} $ & $41.5-169$        & \multirow{2}{*}{LIGO and Virgo}  &  $1.91 \times 10^{-23}$ & $9.08 \times 10^{-26}$ & \multirow{2}{*}{\cite{Aasi:2014zwg}}   \\
$1.8\times10^{-4} $ & $170-600$        &  &  $2.95 \times 10^{-23}$ & $3.83 \times 10^{-26}$ &   \\
$1.2\times10^{5}$ & $0.1 - 1$        & Apollo Seismic   &  $2.93 \times 10^{-16}$ & $1.47 \times 10^{-16}$ & \cite{Coughlin:2014boa} \\
$10^{-15}$ & $0.1 - 1$        & Planck   &  $2.68 \times 10^{-26}$ & $1.34 \times 10^{-26}$ & \cite{Adam:2014bub} \\
\hline                     
\end{tabular}%
}
\caption{Various observational upper limits on $\Omega_{\text{gw}}$ of the stochastic GWs. {The LIGO , Virgo results at different frequencies give different values of $\Omega_{\text{gw}}(f)$ and thus $h_c$ are different.}}
\label{tab2}
\end{table}


\subsubsection{Particle physics application}
\label{subsub322}

Due to our detailed calculation for the KK reduction with the finite size of the extra dimension, one can also estimate the size of the extra dimension by using the results of particle physics observations. From the definition of masses of KK modes, one obtains $L = \frac{I \hbar}{m_{I} c}$. We can find $L$ from the stable radion mass bound. In our model, $h^{I}$ in \eqref{gafix2} corresponds to this radion. Given a stabilization mechanism, the radion mass can in principle be calculated. The required ansatz is that the radion fluctuation about the RS background solves the linearized Einstein equations. Then we incorporate the backreaction of the bulk scalar vacuum expectation value into the metric. Treating the backreaction as a perturbation about the RS solution, the mass bound was given by~\cite{Csaki:2000zn,Tanaka:2000er,Frank:2016oqi}
\begin{align}{\cal O} (10 \text{GeV}) \leq m_{I} \leq {\cal O} (1 \text{TeV}) \, . \label{mr} \end{align} 
Thus, if we adopt this constraint to the mass-length relation in \eqref{dis-rel}, then we obtain
\begin{align} 10^{-19}m \leq L \leq 10 ^{-17}m. \label{sizeL} \end{align} 

\section{Conclusions}
\label{sec:con}

In order to examine the effect of the extra dimension in our observational 4-dimensional phenomena, one needs to consider the small but finite size of the extra dimensions. The conventional KK reduction, where only the massless KK zero modes are kept to obtain 4-dimensional gravity theory, fails to provide evidence on higher-dimensional gravity models.  In the limit of vanishing size of extra dimensions, neglecting the massive KK modes is feasible, whereas, for the small but finite size of the extra dimensions, the massive KK modes can have measurable effects on the 4-dimensional gravity theory. In this manuscript, we investigate the effective 4-dimensional gravity theory obtained from the KK reduction of the Einstein-Hilbert action without the cosmological constant in a compactified 5-dimensional manifold. 

In our model, the KK reduction produces sets of both massless and massive dynamical fields. The massless fields are composed of a tensor mode and a vector mode each with two dynamical degrees of freedom, and a scalar mode with one dynamical degree of freedom. The massive fields also have five degrees of freedom and we chose four massive tensor modes and one massive scalar mode. 
The massive modes act as the matter field and become the sources of the massless fields in the GW equation. In particular, the energy-momentum tensor in the equation of motion of effective 4-dimensional massless graviton is composed of the massive tensor and scalar KK modes. 

In \cite{Andriot:2017oaz}, they claim that there exists one longitudinal massive graviton mode in the KK reduction of 5-dimensional gravity theory. Their model is quite similar to ours, however, we do not find any massive graviton mode in our model.  We found that this discrepancy stems from including $\Delta_{{\cal M}} h_{\mu\nu}$-term in their Eq.(2.22a), which is actually vanishing for the massless mode. Thus, the 4-dimensional effective graviton is massless as we have shown in section \ref{sec:DR}. Our result is also consistent with the previous works \cite{Scherk_ea79,1984PhRvL..52...14D,1992PhRvD..46.2290C,1992PhRvD..46.3483C}.

From the observational upper limit on the size of the extra dimension, we will be able to obtain a lower limit on the frequency of the GW. Higher minimum frequency is required to probe the very small size of the extra dimension.  Also from the various stochastic background GWs  observations, we estimate the amplitude of the massive KK modes $\tilde{h}_{\rho\sigma} \sim 10^{-31} \text{Hz}^{-1}$. Therefore, we propose that high frequency and very sensitive experiments of GWs are required in order to investigate the information on the 5-dimensional compactified extra dimension. However, this high sensitivity might be improved if one considers higher extra dimensional models. As a side remark, we also obtain a bound on the size of the extra dimension as $10^{-19}m \leq L \leq 10^{-17} m$, from the mass bounds on the KK modes. If the observations of the particle physics provide more stringent bound on the mass of the KK modes, it can narrow the frequency searching window.

 
\section*{Acknowledgements}
We would like to thank Hyun Seok Yang and Sang-Heon Yi for
helpful discussions. OK acknowledges the hospitality at APCTP
during the program ``100+4 General Relativity and Beyond'', where part of this work was done. OK, SL, and DT are supported by Basic Science Research Program through the National Research Foun- dation of Korea (NRF) funded by the Ministry of Science, ICT and Future Planning (Grant No. NRF-2017R1D1A1A09000951 (O.K.), NRF- 2017R1A2B4011168 (S.L.) and NRF-2017R1D1A1B03032523 (D.T.), respectively). 

\appendix
\section{Decomposition of Quadratic Order Equations and Harmonic Expansions}\label{QdEq}

We decompose the equations in \eqref{QdFl1} in to the $(\mu,\nu)$, $(\mu,5)$, and $(5,5)$-components. As a result we obtain the following set of equations.
\\

\noindent
$\bullet$ {\bf $(\mu,\nu)$-components:}

\noindent
Choosing the free indices $(p,q)$ in
\eqref{QdFl1} to be $(\mu,\nu)$, we obtain
\begin{align}\label{QdEqmn}
&\nabla^\rho\nabla_\mu h_{\rho\nu}+\nabla^\rho\nabla_\nu h_{\rho\mu}+\nabla^a\nabla_\mu h_{a\nu}+\nabla^a\nabla_\nu h_{a\mu}-(\nabla^\rho_\rho+\nabla^a_a)h_{\mu\nu}-\nabla_\mu\nabla_\nu (h^\rho_\rho+h^a_a)\nn\\
&-g_{\mu\nu}\Big(\nabla^\rho\nabla^\sigma h_{\rho \sigma}+2\nabla^\rho\nabla^ah_{\rho a}+\nabla^a\nabla^bh_{ab}-(\nabla^\rho\nabla_\rho+\nabla^a\nabla_a)(h^\sigma_\sigma+h^b_b)\Big)+Q_{\mu\nu}=0\nn \, , \\ 
\end{align}
where $Q_{\mu\nu}$ is the quadratic part in \eqref{QdTms}. 
Since we are considering a flat background metric, we note that  $\nabla_\rho\nabla_\sigma = \nabla_\sigma\nabla_\rho$. In order to obtain the equation of motion for the tensor zero mode $h^0_{\mu\nu}$, we expand the fluctuation $h_{pq}$ in \eqref{QdEqmn} in terms of the spherical harmonic in \eqref{SHexp} and  project the resulting equations on $Y^0$. Using the gauge fixing \eqref{gafix2}, we obtain
\begin{align}\label{QdEqmn1}
&\nabla^\rho\nabla_\mu h^0_{\rho\nu}+\nabla^\rho\nabla_\nu h^0_{\rho\mu}-\square h^0_{\mu\nu}-\nabla_\mu\nabla_\nu(h^0+ s^0)-g_{\mu\nu}\big(\nabla^\rho\nabla^\sigma h^0_{\rho\sigma}-\square(h^0+ s^0)\big)+ Q_{\mu\nu}^{0} = 0 \, , 
\end{align}

 where
 \begin{align}\label{Qhmn0}
&Q_{\mu\nu}^{0} = -\frac{1}2\nabla_\rho h^{I\rho\sigma}\big(\nabla_\mu h^I_{\sigma\nu}
+\nabla_\nu h^I_{\sigma\mu}\big)+\frac14\nabla_\mu h^{I\rho\sigma}\nabla_\nu h^I_{\rho\sigma}+\frac12\nabla_\rho h^{I\rho\sigma}\nabla_\sigma h^I_{\mu\nu}-\frac12\nabla^\sigma h^I_{\rho\nu}\nabla^\rho h^I_{\sigma\mu}\nn\\
&+\frac12\nabla^\sigma h^{I\rho}_{\nu}\nabla_\sigma h^I_{\mu\rho}-\frac12h^{I\rho\sigma}\nabla_\rho\nabla_\mu h^I_{\sigma\nu}-\frac12h^{I\rho\sigma}\nabla_\rho\nabla_\nu h^I_{\sigma\mu}+\frac12 h^{I\rho\sigma}\nabla_\nu \nabla_\mu h^I_{\rho\sigma}\nn\\
&-\frac{\Lambda^I}2 h^{I\rho}_{\nu}h^I_{\mu\rho}+\frac14\nabla^\rho h^I\big(\nabla_\mu h^I_{\rho\nu}+\nabla_\nu h^I_{\rho\mu}\big)-\frac14\nabla^\rho h^I\nabla_\rho h^I_{\mu\nu}+\frac12h_{\mu\nu}^I\square h^I+\frac{3\Lambda^I}4h^I h^I_{\mu\nu}
\nn \\
& -\frac14\nabla^\rho h^I\nabla_\rho h^I_{\mu\nu}+ \frac14\nabla^\rho h^I\big(\nabla_\mu h^I_{\rho\nu}+\nabla_\nu h^I_{\rho\mu}\big)+ \frac12h_{\mu\nu}^I\square h^I+ \frac{\Lambda^I}4 h^I_{\mu\nu}h^I\nn\\
&+\frac1{4}\nabla_\mu h^I\nabla_\nu {h^I+\frac1{2} h^I \nabla_\mu\nabla_\nu h^I}
+\frac12h^I_{\rho\sigma}\nabla^\rho\nabla^\sigma h^I_{\mu\nu}-\frac12\nabla^\rho\nabla^\sigma h^I_{\rho\sigma} h^I_{\mu\nu}\nn\\
&+g_{\mu\nu}\Big[ \frac12 \nabla_\rho h^{I\rho\sigma}\nabla^\tau h^I_{\sigma\tau}-\frac38 \nabla^\tau h^{I\rho\sigma}\nabla_\tau h^I_{\rho\sigma}+\frac14 \nabla^\sigma h^{I\rho \tau}\nabla_\rho h^I_{\sigma\tau}-\frac12  h^{I\rho\sigma}\square h^I_{\rho\sigma}-\frac{\Lambda^I}8 h^{I\rho\sigma}h^I_{\rho\sigma}
 \nn \\
 &+  h^{I\rho\sigma}\nabla_\rho\nabla^\tau h^I_{\sigma\tau}-\frac12 \nabla_\rho h^{I\rho\sigma}\nabla_\sigma h^I -\frac12  h^{I\rho\sigma}\nabla_\rho\nabla_\sigma h^I+\frac18 \nabla^\rho h^I\nabla_\rho h^I- \frac{\Lambda^I}8 h^Ih^I
\nn \\
&-\frac12 \nabla_\rho h^{I\rho\sigma}\nabla_\sigma h^I-\frac12  h^{I\rho\sigma}\nabla_\rho\nabla_\sigma  h^I+\frac14 \nabla^\rho h^I\nabla_\rho  h^I- \frac{\Lambda^I}4 h^I h^I\nn\\
&-\frac1{4} \nabla^\rho h^I\nabla_\rho h^I-\frac1{2} h^I \square h^I   \Big].
\end{align}
For later convenience, we take the trace of the above equation,
\begin{align}\label{QdEqmn2}
&\square h^0+\frac32 \square s^0-\nabla^\rho\nabla^\sigma h^0_{\rho\sigma}+\frac12 \nabla_\rho h^{I\rho\sigma}\nabla^\tau h^I_{\sigma\tau}-\frac38\nabla^\tau h^{I\rho\sigma}\nabla_\tau h^I_{\rho\sigma}+\frac14\nabla^\sigma h^{I\rho\tau}\nabla_\rho h^I_{\sigma\tau}\nn\\
&+\frac12 h^{I\rho\sigma}\nabla_\rho\nabla^\tau h^I_{\sigma\tau}+ h^{I\rho\sigma}\nabla^\tau\nabla_\rho h^I_{\sigma\tau}-\frac34 h^{I\rho\sigma}\square h^I_{\rho\sigma}-\frac12 {\Lambda^I} h^{I\rho\sigma}h^I_{\rho\sigma}-\frac12 \nabla_\rho h^{I\rho\sigma}\nabla_\sigma h^I
\nn\\
&-\frac34h^{I\rho\sigma}\nabla_\rho \nabla_\sigma h^I-\frac14\nabla^\rho \nabla^\sigma h^I_{\rho\sigma} h^I+\frac18\nabla^\rho h^I\nabla_\rho h^I\boldsymbol+\frac14h^I\square h^I+\frac18\Lambda^Ih^I h^I-\frac34 \nabla_\rho h^{I\rho\sigma}\nabla_\sigma h^I\nn\\
&- h^{I\rho\sigma}\nabla_\rho\nabla_\sigma  h^I+\frac38\nabla^\rho h^I\nabla_\rho h^I+\frac14h^I\square h^I-\frac{3\Lambda^I}8h^I h^I-\frac3{8}\nabla^\rho h^I\nabla_\rho {h^I}-\frac3{4} h^I \square h^I=0.
\end{align}

\noindent
$\bullet$ {\bf $(\mu,5)$-components:} 

\noindent
The $(\mu,a'=5)$-components of \eqref{QdFl1} are given by
\begin{align}\label{QdEqma}
&\nabla^\rho\nabla_\mu h_{\rho{a'}}+\nabla^\rho\nabla_{a'} h_{\rho\mu}+\nabla^a\nabla_\mu h_{a{a'}}+\nabla^a\nabla_{a'} h_{a\mu}-(\nabla^\rho_\rho+\nabla^a_a)h_{\mu{a'}}-\nabla_\mu\nabla_{a'} (h^\rho_\rho+h^a_a)\nn\\
&-g_{\mu{a'}}\Big(\nabla^\rho\nabla^\sigma h_{\rho \sigma}+2\nabla^\rho\nabla^ah_{\rho a}+\nabla^a\nabla^bh_{ab}-(\nabla^\rho\nabla_\rho+\nabla^a\nabla_a)(h^\sigma_\sigma+h^b_b)\Big)+Q_{\mu5}=0.
 \end{align}
Projecting on $Y^0$ and $\nabla^5 Y^I$, we obtain the equation of motion for the vector zero mode and some constraint equations for $h^I_{\mu\nu}$
\begin{align}
&\nabla_\mu\nabla^\nu v^0_\nu-\square v^0_\mu+Q^0_\mu=0,\qquad\qquad \nabla^\nu h^I_{\mu\nu}-\nabla_\mu h^I+Q^I_\mu=0,
\end{align}
where we have used the gauge fixing in \eqref{gafix2}, and $(Q^0_\mu, Q^I_\mu)$ are quadratic terms, whose explicit forms are not needed for what we want to accomplish in this paper.\\

\noindent
$\bullet$ {\bf $(5,5)$-components:}

\noindent
The $(a'=5,b'=5)$-components of   
\eqref{QdFl1} are 
\begin{align}\label{QdEqaa}
&\nabla^\rho\nabla_{a'} h_{\rho{b'}}+\nabla^\rho\nabla_{b'} h_{\rho{a'}}+\nabla^a\nabla_{a'} h_{a{b'}}+\nabla^a\nabla_{b'} h_{a{a'}}-(\nabla^\rho_\rho+\nabla^a_a)h_{{a'}{b'}}-\nabla_{a'}\nabla_{b'} (h^\rho_\rho+h^a_a)\nn\\
&-g_{{a'}{b'}}\Big(\nabla^\rho\nabla^\sigma h_{\rho \sigma}+2\nabla^\rho\nabla^ah_{\rho a}+\nabla^a\nabla^bh_{ab}-(\nabla^\rho\nabla_\rho+\nabla^a\nabla_a)(h^\sigma_\sigma+h^b_b)\Big)+Q_{a'b'} =0.
\end{align}
Again we expand \eqref{QdEqaa} in terms of the spherical harmonic \eqref{SHexp} and project the resulting equation  on $Y^0$. With our gauge fixing condition \eqref{gafix2}, we obtain
\begin{align}\label{QdEqh0}
&\square h^0-\nabla^\rho\nabla^\sigma h^0_{\rho\sigma}+ Q_{h}^{0} = 0, 
\end{align}
where 
\begin{align}\label{Qh0}
&Q_{h}^{0} = \frac12\nabla_\rho h^{I\rho\sigma}\nabla^\tau h^I_{\sigma\tau}-\frac{3}8\nabla^\tau h^{I\rho\sigma}\nabla_\tau h^I_{\rho\sigma}+\frac14\nabla^\sigma h^{I\rho \tau}\nabla_\rho h^I_{\sigma\tau}
\nn \\
&+h^{I\rho\sigma}\nabla_\rho\nabla^\tau h^I_{\sigma\tau}-\frac12 h^{I\rho\sigma}\square h^I_{\rho\sigma}+\frac{\Lambda^I}8 h^{I\rho\sigma}h^I_{\rho\sigma}-\frac12\nabla_\rho h^{I\rho\sigma}\nabla_\sigma h^I-\frac12 h^{I\rho\sigma}\nabla_\rho\nabla_\sigma h^I
\nn\\
&+\frac18\nabla^\rho h^I\nabla_\rho h^I-\frac{\Lambda^I}8 h^Ih^I-\frac12  h^I\nabla^\rho\nabla^\sigma h^I_{\rho\sigma }+\frac12  h^I\square h^I.
\end{align}

Combining \eqref{QdEqmn2} and \eqref{QdEqh0}, we obtain the  equation of motion for zero modes  $s^0$ 
\begin{align}\label{QdEqs0}
& \square s^0 + Q_{s}^{0} = 0,
\end{align}
where
\begin{align}\label{Qs0}
& Q_{s}^{0} \equiv - \Bigg( - \frac13 h^{I\rho\sigma}\nabla_\rho\nabla^\tau h^I_{\sigma\tau} +\frac16  h^{I\rho\sigma}\square h^I_{\rho\sigma}+\frac{5\Lambda^I}{12} h^{I\rho\sigma}h^I_{\rho\sigma}+ \frac12 \nabla^\rho h^I\nabla^\sigma h^I_{\rho \sigma}  
\nn\\
& +\frac16  h^I\nabla^\rho \nabla^\sigma h^I_{\rho\sigma}+\frac16 h^{I\rho\sigma}\nabla_\rho \nabla_\sigma h^I-\frac16 h^I\square h^I- \frac1{6}\Lambda^I h^Ih^I 
\nn \\
& +\frac23 h^{I\rho\sigma}\nabla_\rho\nabla_\sigma h^I-\frac13   h^I\nabla^\rho\nabla^\sigma h^I_{\rho\sigma }-\frac14 \nabla^\rho h^I \nabla_\rho h^I - \frac16 h^I\square h^I+\frac{\Lambda^I}4   h^I h^I \nn\\
&+ \frac{1}{4}\nabla^\rho  h^I\nabla_\rho  h^I +\frac13 h^I\square h^I+\frac12  h^I \square h^I \bigg).
\end{align}

Inserting \eqref{QdEqh0} and \eqref{QdEqs0} into \eqref{QdEqmn1}, we obtain 
\begin{align}\label{QdEqhmn}
\square h^0_{\mu\nu}=&\nabla^\rho\nabla_\mu h^0_{\rho\nu}+\nabla^\rho\nabla_\nu h^0_{\rho\mu}-\nabla_\mu\nabla_\nu(h^0+ s^0)\nn\\
&-\frac{1}2\nabla_\rho h^{I\rho\sigma}\big(\nabla_\mu h^I_{\sigma\nu}
+\nabla_\nu h^I_{\sigma\mu}\big)+\frac12\nabla_\rho h^{I\rho\sigma}\nabla_\sigma h^I_{\mu\nu}+\frac14\nabla_\mu h^{I\rho\sigma}\nabla_\nu h^I_{\rho\sigma}
\nn \\
&-\frac12\nabla^\sigma h^I_{\rho\nu}\nabla^\rho h^I_{\sigma\mu}+\frac12\nabla^\sigma h^{I\rho}_{\nu}\nabla_\sigma h^I_{\mu\rho}+\frac12h^{I\rho\sigma}\nabla_\rho \nabla_\sigma h^I_{\mu\nu}-\frac12\nabla^\rho \nabla^\sigma h^I_{\rho\sigma} h^I_{\mu\nu}
\nn \\
&-\frac12h^{I\rho\sigma}\nabla_\rho\nabla_\mu h^I_{\sigma\nu}-\frac12h^{I\rho\sigma}\nabla_\rho\nabla_\nu h^I_{\sigma\mu}+\frac12 h^{I\rho\sigma}\nabla_\nu \nabla_\mu h^I_{\rho\sigma}\nn\\
&+\frac14\nabla^\rho h^I\big(\nabla_\mu h^I_{\rho\nu}+\nabla_\nu h^I_{\rho\mu}\big)-\frac{\Lambda^I}2 h^{I\rho}_{\nu}h^I_{\mu\rho}-\frac14\nabla^\rho h^I\nabla_\rho h^I_{\mu\nu}+\frac12h^I_{\mu\nu}\square h^I\nn\\
&+\frac{3\Lambda^I}4h^I h^I_{\mu\nu}+ \frac14\nabla^\rho h^I\big(\nabla_\mu h^I_{\rho\nu}+\nabla_\nu h^I_{\rho\mu}\big)-\frac14\nabla_\rho h^I_{\mu\nu}\nabla^\rho h^I+\frac12h^I_{\mu\nu}\square h^I
\nn \\
& + \frac{\Lambda^I}4 h^I_{\mu\nu}h^I+\frac1{4}\nabla_\mu h^I\nabla_\nu {h^I+\frac1{2} h^I \nabla_\mu\nabla_\nu h^I}
\nn\\
&+g_{\mu\nu}\Big(\frac{1}{6}h^{I\rho\sigma}\square h^I_{\rho\sigma}-\frac{1}{3}h^{I\rho\sigma}\nabla_\rho\nabla^\tau h^I_{\sigma\tau}+\frac{\Lambda^I}6h^{I\rho\sigma}h^I_{\rho\sigma}+\frac{1}{6}h^I\nabla^\rho \nabla^\sigma h^I_{\rho\sigma}
\nn \\
&  +\frac{1}{6}h^{I\rho\sigma}\nabla_\rho\nabla_\sigma h^I-\frac16h^I\square h^I-\frac{\Lambda^I}6 h^Ih^I+\frac16h^{I\rho\sigma}\nabla_\rho\nabla_\sigma  h^I+\frac{1}{6}h^I\nabla^\rho\nabla^\sigma h^I_{\rho\sigma }
\nn \\
& -\frac16h^I\square h^I-\frac{1}{6}h^I\square h^I\Big).
\end{align}
Combining \eqref{QdEqh0}, \eqref{QdEqs0}, and \eqref{QdEqhmn}, we have the relation
\begin{align}\label{QdEqhmn1}
&L_E\hat h^0_{\mu\nu}-\frac{1}{4}\nabla_\rho h^{I\rho\sigma}\nabla_\mu h^I_{\sigma\nu}
-\frac{1}{4}\nabla_\rho h^{I\rho\sigma}\nabla_\nu h^I_{\sigma\mu}+\frac{1}{4}\nabla_\rho h^{I\rho\sigma}\nabla_\sigma h^I_{\mu\nu}+\frac{1}{8}\nabla_\mu h^{I\rho\sigma}\nabla_\nu h^I_{\rho\sigma}
\nn \\
&-\frac{1}{4}\nabla^\sigma h^I_{\rho\nu}\nabla^\rho h^I_{\sigma\mu}+\frac{1}{4}\nabla^\sigma h^{I\rho}_{\nu}\nabla_\sigma h^I_{\mu\rho}-\frac{1}{4}\nabla^\rho \nabla^\sigma h^I_{\rho\sigma} h^I_{\mu\nu}-\frac{1}{4}h^{I\rho\sigma}\nabla_\rho\nabla_\mu h^I_{\sigma\nu}\nn\\
&-\frac{1}{4}h^{I\rho\sigma}\nabla_\rho\nabla_\nu h^I_{\sigma\mu}+\frac{1}{4}h^{I\rho\sigma}\nabla_\rho \nabla_\sigma h^I_{\mu\nu}+\frac{1}{4}h^{I\rho\sigma}\nabla_\nu \nabla_\mu h^I_{\rho\sigma}-\frac{\Lambda^I}4 h^I_{\mu\rho}h^{I\rho}_{\nu}
\nn \\
& -\frac{1}{8}\nabla_\rho h^I_{\mu\nu}\nabla^\rho h^I+\frac{1}{8}\big(\nabla_\mu h^I_{\rho\nu}+\nabla_\nu h^I_{\rho\mu}\big)\nabla^\rho h^I+\frac14h^I_{\mu\nu}\square h^I+\frac{3\Lambda^I}{8}h^I_{\mu\nu}h^I
\nn\\
&+\frac{1}{8}\big(\nabla_\mu h^I_{\rho\nu}+\nabla_\nu h^I_{\rho\mu}\big)\nabla^\rho h^I-\frac{1}{8}\nabla_\rho h^I_{\mu\nu}\nabla^\rho h^I +\frac14h^I_{\mu\nu}\square h^I+ \frac{\Lambda^I}{8} h^I_{\mu\nu}h^I
\nn \\
& +\frac{1}{8}\nabla_\mu h^I\nabla_\nu {h^I+\frac{1}{4}h^I \nabla_\mu\nabla_\nu h^I}
\nn\\
&+ g_{\mu\nu} \Big(-\frac{1}{4} h^{I\rho\sigma}\nabla^\tau\nabla_\rho h^I_{\tau\sigma}+\frac{1}{8} h^{I\rho\sigma}\square h^I_{\rho\sigma}+\frac{3\Lambda^I}{16} h^{I\rho\sigma}h^I_{\rho\sigma} +\frac18  h^{I\rho\sigma}\nabla_\rho\nabla_\sigma h^I
\nn \\
&+\frac{1}{8}\nabla^\rho \nabla^\sigma h^I_{\rho\sigma} h^I-\frac{1}{8} h^I\square h^I- \frac{\Lambda^I}{8}h^Ih^I+\frac18 \nabla_\rho h^{I\rho\sigma}\nabla_\sigma h^I+\frac14 h^{I\rho\sigma}\nabla_\rho\nabla_\sigma h^I\nn\\
& -\frac{1}{16} \nabla^\rho h^I\nabla_\rho h^I-\frac{1}{8} h^I\square h^I+\frac{\Lambda^I}{16}h^I h^I+\frac{1}{16} \nabla^\rho h^I\nabla_\rho h^I+\frac{1}{8} h^I \square h^I  \Big)=0,
\end{align}
where $\hat h^0_{\mu\nu}=h^0_{\mu\nu}+\frac12 g_{\mu\nu}s^0$ is the effective 4-dimensional graviton.

\section{Effective Energy-momentum Tensor from GW perturbation}
\label{sec:AppB}

In this appendix, we briefly review the four dimensional energy-momentum tensor from GW perturbation, in order to show the analogy with what we done in the section \ref{sec:three}. In the section \ref{sec:DR}, starting from the flat 5-dimensional gravity theory and applying the KK reduction including terms that are quadratic order in the metric perturbation, we have obtained the 4-dimensional effective gravity theory with matter. The similar situation has been considered in the 4-dimensional gravity when a source of GW is far from the observer. In this case, the energy carried by a wave as it leaves a source can be written as an effective stress-energy tensor for the wave by including the second order perturbation. This is the so-called Isaacson energy-momentum tensor~\cite{Isaacson}. 

We assume that the amount of energy associated with GW in a region of spacetime  is large enough to contain several wavelengths of gravitational radiation but is smaller than  any background curvature scale. In this case, one need to do an integral average of an effective energy-momentum tensor over a volume large enough that bulk contributions are greater than the boundary contributions.  Einstein field equations in a vacuum are written by
\begin{align} G_{\mu\nu} &= \overset{0}{G}_{\mu\nu} + \overset{1}{G}_{\mu\nu}[h_{\alpha\beta}] + \overset{2}{G}_{\mu\nu}[h_{\alpha\beta}] + \cdots \nonumber \\
&= 0  \,  . \label{GmunuO} \end{align} 
One can write the metric in the flat spacetime with its fluctuation as
\begin{align} g_{\alpha\beta} &= \eta_{\alpha\beta} + h_{\alpha\beta} =  \eta_{\alpha\beta} + \lambda \overset{1}{h}_{\alpha\beta}  + \lambda^2 \overset{2}{h}_{\alpha\beta} + \cdots \,\, , \,\,\,\,\,\,\,\,\,\,\, |h_{\alpha\beta}| \ll 1, \label{getah} \end{align}  
where $\lambda$ is a formal order parameter denoting the first order correction ($\overset{1}{h}$) and the second order correction ($\overset{2}{h}$), respectively. Then one can rewrite (\ref{GmunuO}) by using (\ref{getah})
as\begin{align} G_{\mu\nu} &= \overset{0}{G}_{\mu\nu} + \lambda \overset{1}{G}_{\mu\nu}[\overset{1}{h}_{\alpha\beta}] + \lambda^2 \left( \overset{1}{G}_{\mu\nu}[\overset{2}{h}_{\alpha\beta}] + \overset{2}{G}_{\mu\nu}[\overset{1}{h}_{\alpha\beta}]  \right) + \mathcal{O} (\lambda^3)  + \cdots \nonumber \\
&= 0.  \end{align} 
The above equation should hold order-by-order in $\lambda$, then the first-order and second-order corrections to the background metric are given by \cite{MTW}
\begin{align} \overset{0}{G}_{\mu\nu} &= 0, \label{O0} \\
\overset{1}{G}_{\mu\nu} [ \overset{1}{h}_{\alpha\beta} ] &= 0, \label{O1} \\
\overset{1}{G}_{\mu\nu} [ \overset{2}{h}_{\alpha\beta} ] &= -\overset{2}{G}_{\mu\nu} [ \overset{1}{h}_{\alpha\beta} ] = \frac{8 \pi G}{c^4} T_{\mu\nu}^{\text{(GW)}}, \label{O2} \end{align} 
where $T_{\mu\nu}^{\text{(GW)}} \equiv - \frac{c^4}{8 \pi G} \overset{2}{G}_{\mu\nu} [ \overset{1}{h}_{\alpha\beta} ]$ is the effective energy-momentum tensor created by the 1st order GW perturbation. 
 In order to make $T_{\mu\nu}^{\text{(GW)}}$ gauge invariant, one must perform an integral average over a region of spacetime large enough to contain several GW oscillations
\begin{align} T_{\mu\nu}^{\text{GW}} &= - \frac{c^4}{8 \pi G} \left< \overset{2}{R}_{\mu\nu} - \frac{1}{2} \eta_{\mu\nu} \overset{2}{R} \right> \, , \label{TGWO2} \end{align} 
where $<>$ represents the integral average. To perform the above calculation, one uses the definition of Ricci tensor, $R_{\mu\nu} = \frac{\partial \Gamma^{\alpha}_{\mu\nu}}{\partial x^{\alpha}} - \frac{\partial \Gamma^{\alpha}_{\alpha\nu}}{\partial x^{\mu}} + \Gamma^{\alpha}_{\mu\nu} \Gamma^{\beta}_{\alpha\beta} - \Gamma^{\alpha}_{\beta\nu} \Gamma^{\beta}_{\mu\alpha}$
and retain the terms quadratic in $h_{\mu\nu}$ while applying the following tricks : i) adopt TT-gauge so that $h_{0\mu}^{\text{TT}}=0$, $\delta^{ik} \partial h_{ij}^{\text{TT}} / \partial x^{k}=0$, and $\delta^{ij}h_{ij}^{\text{TT}}=0$. ii) assume that under integral average, all terms of the form $<\partial T_{\mu \cdots \gamma} / \partial x^{\nu}>$ can be neglected since such terms contribute only to the boundary of the region and can be made arbitrarily small compared to the bulk by expanding the region in which the integral average is performed. iii) use $\Box h_{ij}^{\text{TT}}=0$.


\end{document}